\documentclass{article}

    \PassOptionsToPackage{numbers, compress}{natbib}

    \usepackage[final]{neurips_2020}

\usepackage[utf8]{inputenc} 
\usepackage[T1]{fontenc}    
\usepackage{hyperref}       
\usepackage{url}            
\usepackage{booktabs}       
\usepackage{amsfonts}       
\usepackage{nicefrac}       
\usepackage{microtype}      
\usepackage{amsmath}
\usepackage{natbib}
\usepackage{algorithm}
\usepackage{booktabs} 
\usepackage{caption}
\usepackage{xcolor}
\usepackage[noend]{algpseudocode}
\usepackage{wrapfig}
\usepackage{adjustbox}
\usepackage{subfigure}
\usepackage{changepage}
\usepackage{sidecap}

\def\eg{\emph{e.g.~}}

\def\ie{\emph{i.e.~}}

\title{Why are we baffled by adversarial examples!?}
\title{The real reason behind adversarial examples!}
\title{Harnessing adversarial examples with \\ a surprisingly simple defense}

\author{
Ali Borji \\
\texttt{aliborji@gmail.com} 
\thanks{Code is available at: \url{https://github.com/aliborji/ReLU_defense.git}}
}

\begin{document}

\maketitle

\begin{abstract}
    I introduce a very simple method to defend against adversarial examples. The basic idea is to raise the slope of the ReLU function at the test time. Experiments over MNIST and CIFAR-10 datasets demonstrate the effectiveness of the proposed defense against a number of strong attacks in both untargeted and targeted settings. While perhaps not as effective as the state of the art adversarial defenses, this approach can provide insights to understand and mitigate adversarial attacks. It can also be used in conjunction with other defenses.
\end{abstract}



\section{Introduction}
\begin{figure}[h]       
    \vspace{-10px}
    \centering
    \includegraphics[width=.5\linewidth]{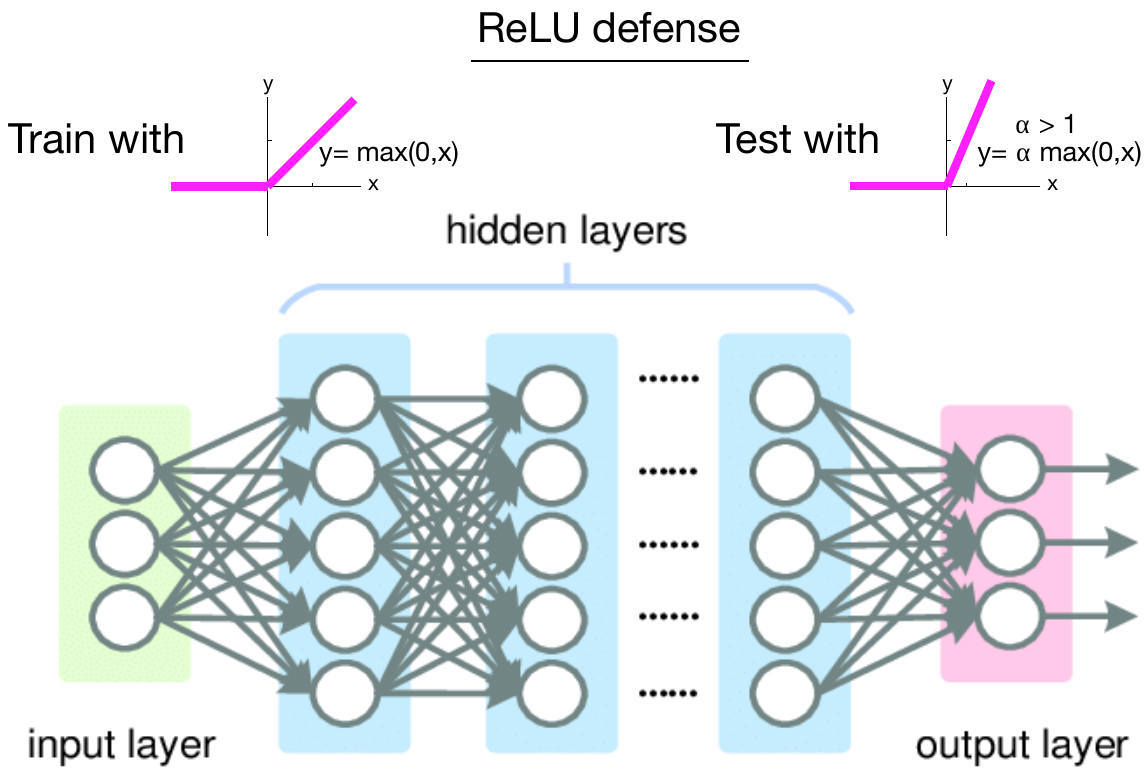}
    \caption{Sketch of the the proposed defense. Just increase the ReLU slope at the test time!}
    \label{fig:idea}
\end{figure}

\begin{figure}[h]       
\begin{center}
\includegraphics[width=\linewidth]{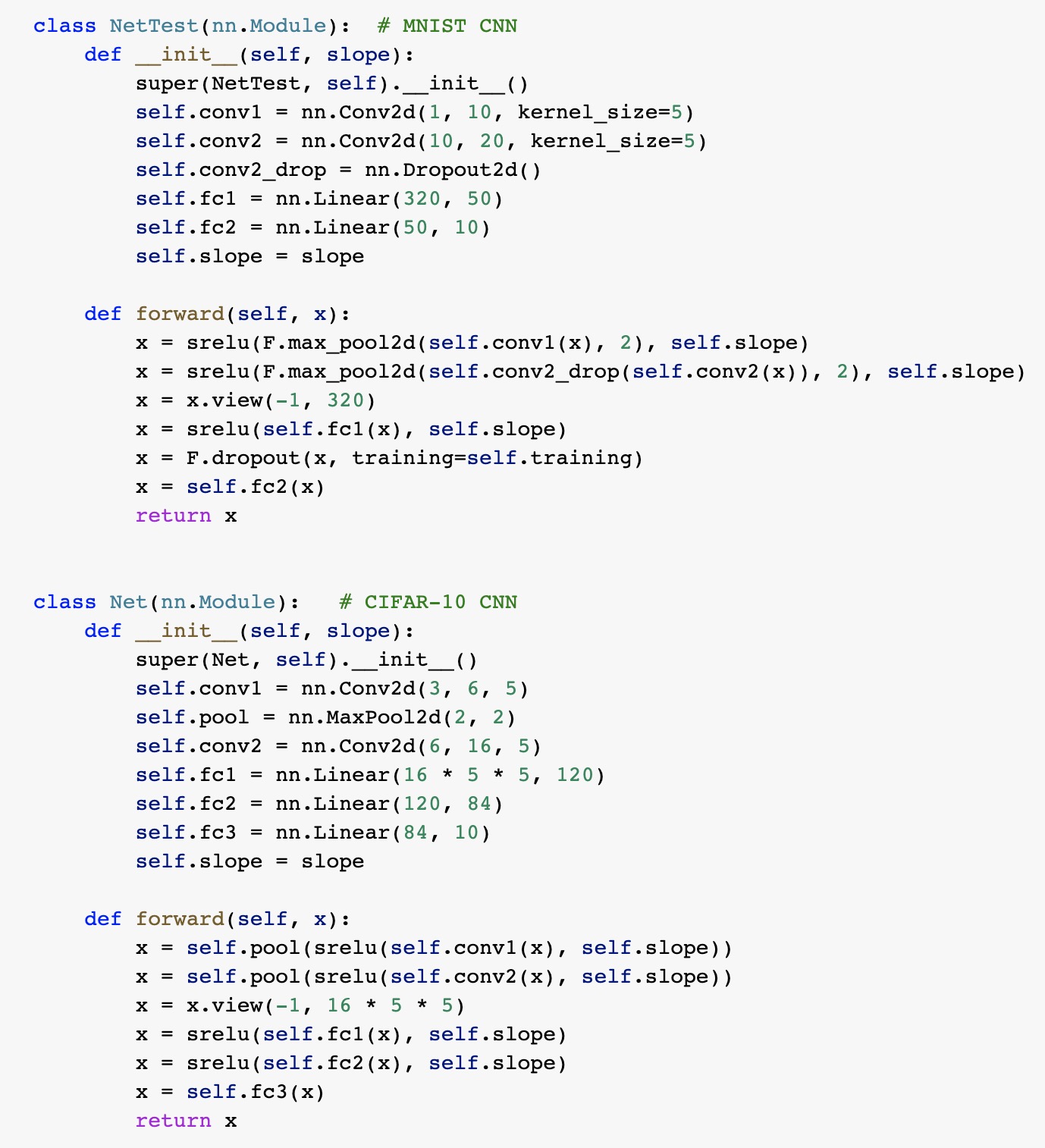} 
\caption{CNN architectures used in the experiments (top: MNIST, bottom: CIFAR10-CNN1).}
\label{fig:appx}
\end{center}
\end{figure}

The lockdown has not been too bad! After all, we found some time to explore what we always wanted to but did not have time for it. For me, I have been passing the time by immersing myself in adversarial ML. I was playing with the adversarial examples tutorial in PyTorch\footnote{\url{https://pytorch.org/tutorials/beginner/fgsm_tutorial.html}} and came across something interesting. So, I decided to share it with you. It is a simple defense that works well against untargeted attacks, and to some extent against targeted ones. Here is how it goes. The idea is to train a CNN with the ReLU activation function but increase its slope at the test time (Fig.~\ref{fig:idea}). Lets call this function Sloped ReLU or SReLU for short: $\text{SReLU}(\alpha,x)= \alpha \ \text{max}(0,x)$, where $\alpha$ is the slope. SReLU becomes ReLU for $\alpha=1$.
To investigate this idea, I ran the following CNNs (Fig.~\ref{fig:appx}) over MNIST~\cite{lecun1998gradient}: 
\centerline{Conv $\Rightarrow$ Pool $\Rightarrow$ SReLU $\Rightarrow$ 
Conv $\Rightarrow$ Pool $\Rightarrow$ SReLU $\Rightarrow$ 
FC $\Rightarrow$ SReLU $\Rightarrow$ FC} 
and over CIFAR-10~\cite{krizhevsky2009learning} (referred to as CIFAR10-CNN1): \\
\centerline{Conv $\Rightarrow$ SReLU $\Rightarrow$ Pool $\Rightarrow$ Conv $\Rightarrow$ SReLU $\Rightarrow$ Pool $\Rightarrow$ FC $\Rightarrow$ SReLU $\Rightarrow$ FC $\Rightarrow$ SReLU $\Rightarrow$ FC} \\
I also tried a variation of the latter network with SReLUs only after the first two FC layers (referred to as CIFAR10-CNN2). I chose $\alpha \in \{0.5, 1, 2, 5, 10, 100\}$. 


\section{Experiments and results}

Here, I emphasize on the FGSM attack since it is very straightforward~\cite{goodfellow2015explaining}. To craft an adversarial example $x_{adv}$, FGSM adds a tiny portion of the gradient of the loss w.r.t the input image back to the input image (\ie gradient ascent in loss):
\begin{equation}
\mathbf{x}_{adv} \leftarrow{} \text{clip}(\mathbf{x} + \epsilon \times sign(\nabla_{\mathbf{x}} J(\mathbf{\theta}, \mathbf{x}, y)))
\end{equation}
The perturbed image needs to be clipped to the right range (here [0,1] in all experiments over both datasets). $\epsilon \in [0,1]$ balances the attack success rate versus imperceptibility. $\epsilon=0$ corresponds to the model performance on the original test set (\ie no perturbation). To gain a higher attack rate more perturbation (\ie larger $\epsilon$) is needed which leads to a more noticeable change (and vice versa). 

The above formulation is for the untargeted attack. For the targeted attack, instead of increasing the loss for the true class label, we can lower the loss for the desired target class label $y_{targ}$ (or we could do both):
\begin{equation}
\mathbf{x}_{adv} \leftarrow \text{clip}(\mathbf{x} - \epsilon \times sign(\nabla_{\mathbf{x}} J(\mathbf{\theta}, \mathbf{x}, y_{targ})))
\end{equation}

In addition to FGSM, I also considered a number of other strong attacks including BIM (also known as IFGSM; iterative FGSM)~\cite{kurakin2016physical}, RFGSM~\cite{tramer2017ensemble}, StepLL~\cite{kurakin2016physical}, PGD-40~\cite{madry2017towards}, and DeepFool~\cite{dezfooli2016deepfool}.
In almost all of these attacks (except DeepFool for which I varied the number of iterations), there is a parameter that controls the magnitude of perturbation (here  represented by $\epsilon$). 
In what follows I will show the results over MNIST and CIFAR-10 datasets against both untargeted and targeted attacks.

\begin{figure}[t]       
    \includegraphics[width=.33\linewidth]{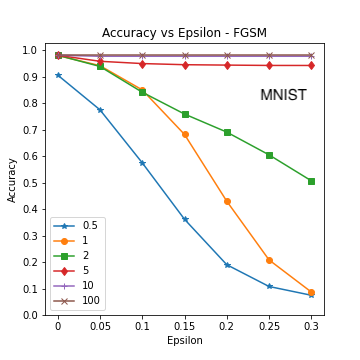}     \vspace{-16px}
    \includegraphics[width=.33\linewidth]{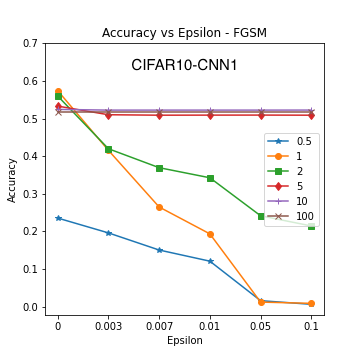}
    \includegraphics[width=.33\linewidth]{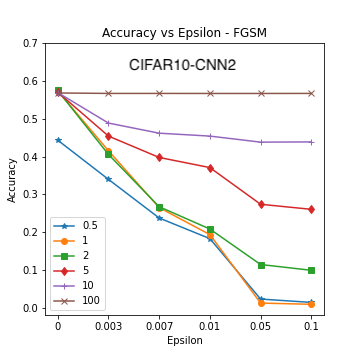} \\

    \includegraphics[width=.33\linewidth]{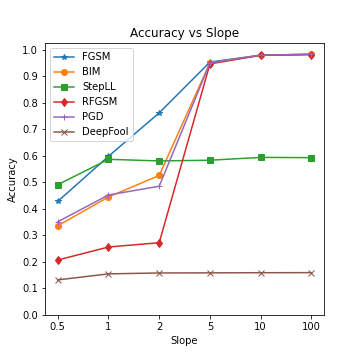} \vspace{-16px}
    \includegraphics[width=.33\linewidth]{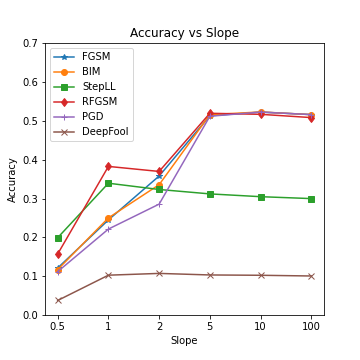}  
    \includegraphics[width=.33\linewidth]{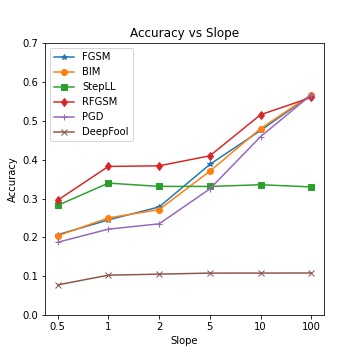}   \\

    \includegraphics[width=.33\linewidth]{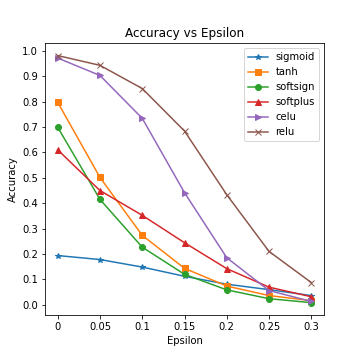}     \includegraphics[width=.33\linewidth]{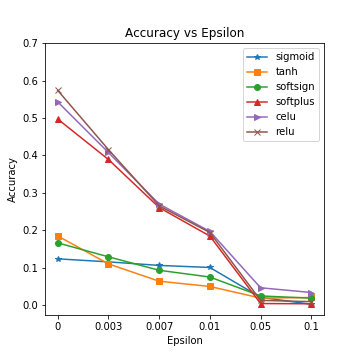}
    \includegraphics[width=.33\linewidth]{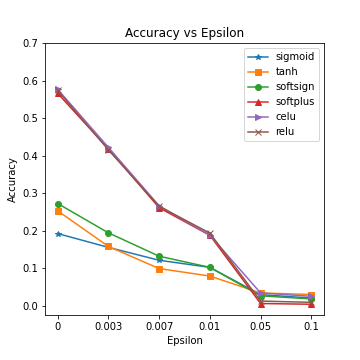}

    \caption{Performance of the ReLU defense against {\bf untargeted attacks}. Left column: MNIST. Middle column: CIFAR-10 with all ReLUs replaced with SReLU, \ie CIFAR10-CNN1. Right column: CIFAR-10 with SReLUs only after the first two FC layers, \ie CIFAR10-CNN2. All 10K test images of each dataset were used.
    Rows from top to bottom: ReLU defense against the FGSM attack, the defense success rate (\ie accuracy loss recovery) as a function of SReLU slope averaged over all epsilons (iterations for DeepFool), and the effect of switching to a different activation function against the FGSM attack. Increasing the SReLU slope improves the performance significantly, except over StepLL and DeepFool attacks. $\epsilon=0$ corresponds to the classifier accuracy with no input perturbation. 
    {\bf Higher y value here means better defense.}}
        
    \label{fig:untargeted}
\end{figure}


\begin{figure}[!h]       
\begin{center}
    
    \vspace{-35pt}    
    \includegraphics[width=.3\linewidth]{./figs/MNIST/MNIST-FGSM.png}
    \hspace{10pt}
    \includegraphics[width=.3\linewidth]{./figs/CIFAR10/CNN1/CIFAR10-FGSM.png}\hspace{10pt}
    \includegraphics[width=.3\linewidth]{./figs/CIFAR10/CNN2/CIFAR10-FGSM.png} \\
    \vspace{-10pt}    
    
    \includegraphics[width=.3\linewidth]{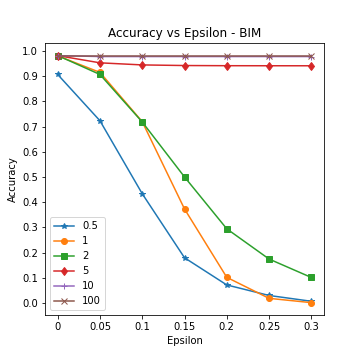}
    \hspace{10pt}
    \includegraphics[width=.3\linewidth]{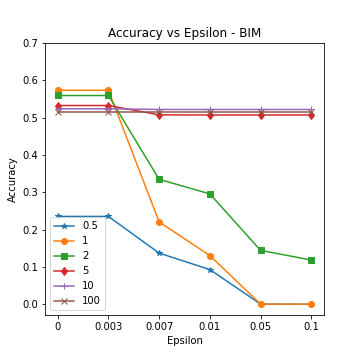}
    \hspace{10pt}
    \includegraphics[width=.3\linewidth]{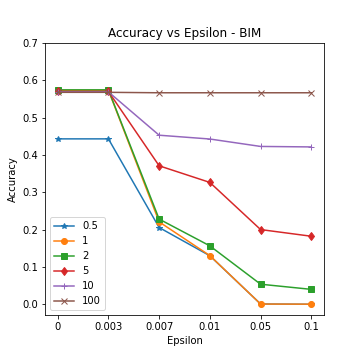} \\
    \vspace{-10pt}

    \includegraphics[width=.3\linewidth]{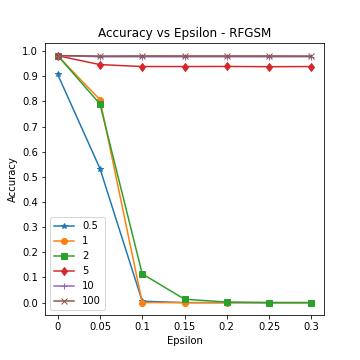} 
    \hspace{10pt}
    \includegraphics[width=.3\linewidth]{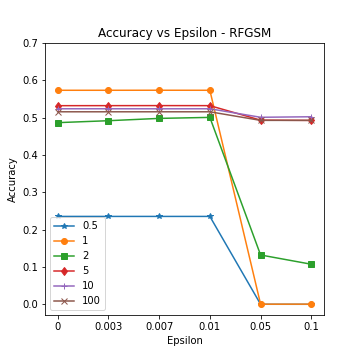} 
    \hspace{10pt}
    \includegraphics[width=.3\linewidth]{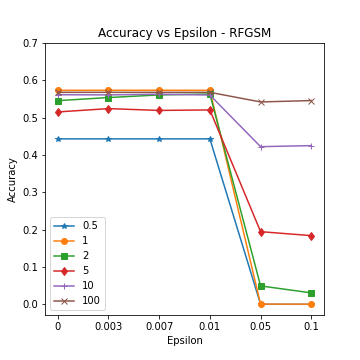}  \\         \vspace{-10pt}

    \includegraphics[width=.3\linewidth]{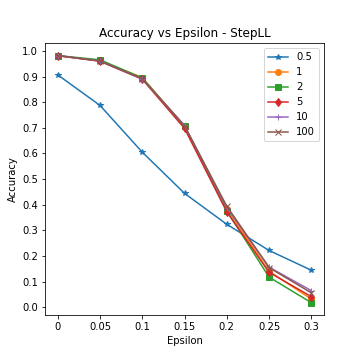}
    \hspace{10pt}
    \includegraphics[width=.3\linewidth]{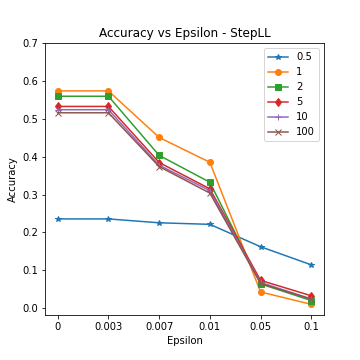}
    \hspace{10pt}
    \includegraphics[width=.3\linewidth]{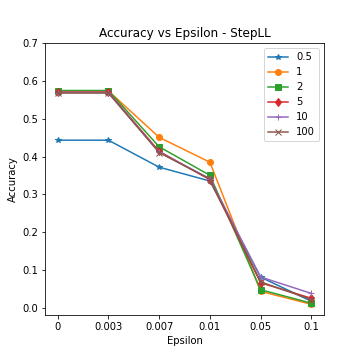} \\         \vspace{-10pt}

    \includegraphics[width=.3\linewidth]{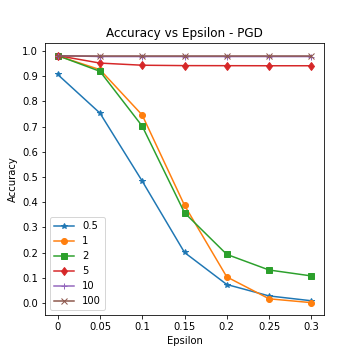}
    \hspace{10pt}
    \includegraphics[width=.3\linewidth]{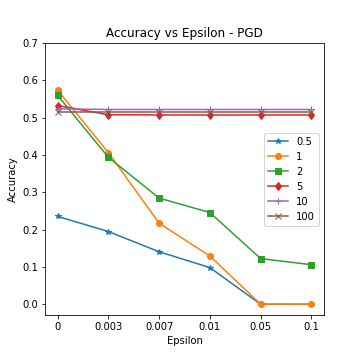}
    \hspace{10pt}
    \includegraphics[width=.3\linewidth]{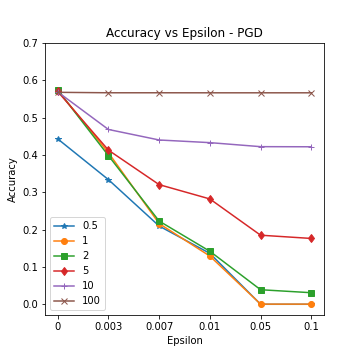} \\
    \vspace{-10pt}

    \includegraphics[width=.3\linewidth]{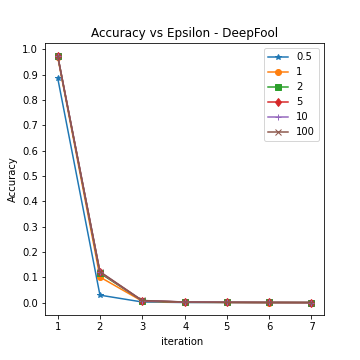} 
    \hspace{10pt}
    \includegraphics[width=.3\linewidth]{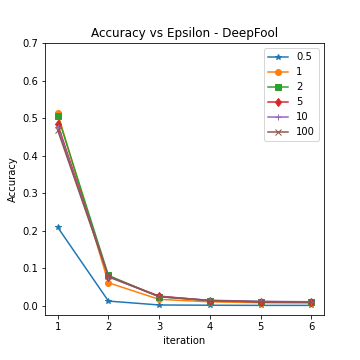}    
    \hspace{10pt}
    \includegraphics[width=.3\linewidth]{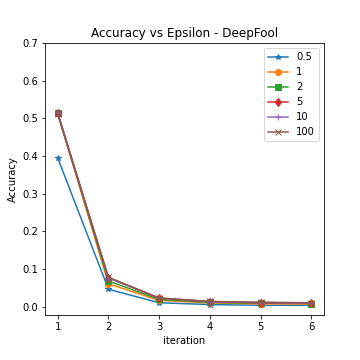}   \\ 
    
    \vspace{-5pt}    
    \caption{Results of the ReLU defense against untargeted attacks over MNIST (left column), CIFAR10-CNN1 (middle), and CIFAR10-CNN2 (right). {\bf Higher y value here means better defense.}}         
    
\label{fig:untargeted_models}    
\end{center}
\end{figure}

\subsection{Performance against untargeted attacks}
Results are shown in Fig.~\ref{fig:untargeted} over entire test sets of datasets. I did not find the SReLU slopes smaller than one very interesting but I am including them here for the sake of comparison. Their effect is not consistent. They are useful against targeted attacks but almost always hinder both classifier accuracy and robustness in untargeted attack settings. 

Over MNIST, at $\alpha=1$ and $\epsilon=0$ (top left panel in Fig.~\ref{fig:untargeted}) classifier accuracy is around 98\% and gradually falls as $\epsilon$ grows. For $\alpha \in \{2,5\}$, the performance loss starts to recover. Finally, for $\alpha \in \{10,100\}$ the classifier has not been impacted at all, hence completely failing the attack. Classifier performance as a function of slope (averaged over epsilons) for each attack type is shown in the middle row of Fig.~\ref{fig:untargeted}. The ReLU defense does very well against FGSM, BIM, RFGSM and PGD attacks. Over StepLL and DeepFool, however, it neither helped nor harmed the classifier.  
Results over the CIFAR-10 dataset are consistent with the MNIST results (middle and right columns in Fig.~\ref{fig:untargeted} corresponding to CIFAR10-CNN1 and CIFAR10-CNN2, respectively). On this dataset, the classifier had around 60\% accuracy on the clean test set ($\epsilon=0$) and went down with more perturbation. Here again increasing $\alpha$ led to better defense, although sometimes at the expense of accuracy. Please see Fig.~\ref{fig:untargeted_models} for the performance of each attack type as a function of perturbation magnitude.

Are these findings specific to SReLU function? To answer this question, I swapped the SReLU with other activation functions and measured the FGSM performance. Now the attacks became even more damaging (Fig.~\ref{fig:untargeted}; bottom row) indicating that SReLU does better.

\begin{figure}[h] 
\begin{adjustwidth}{-2cm}{-2cm}
    \hspace{-45pt}
    \includegraphics[width=.4\linewidth]{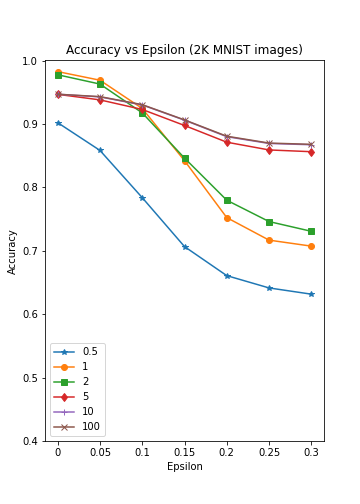}
    \hspace{-15pt}
    \includegraphics[width=.4\linewidth]{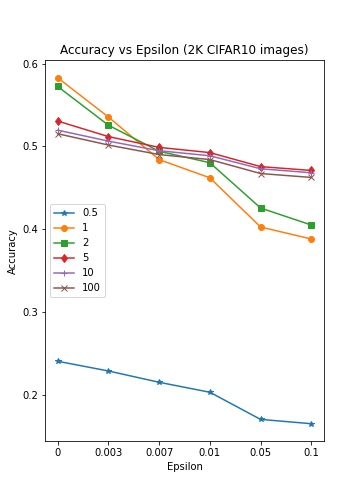} 
    \hspace{-15pt}
    \includegraphics[width=.4\linewidth]{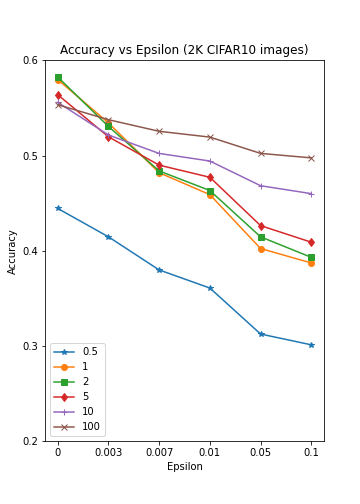} \\
    
    \hspace{-45pt}
    \includegraphics[width=.4\linewidth]{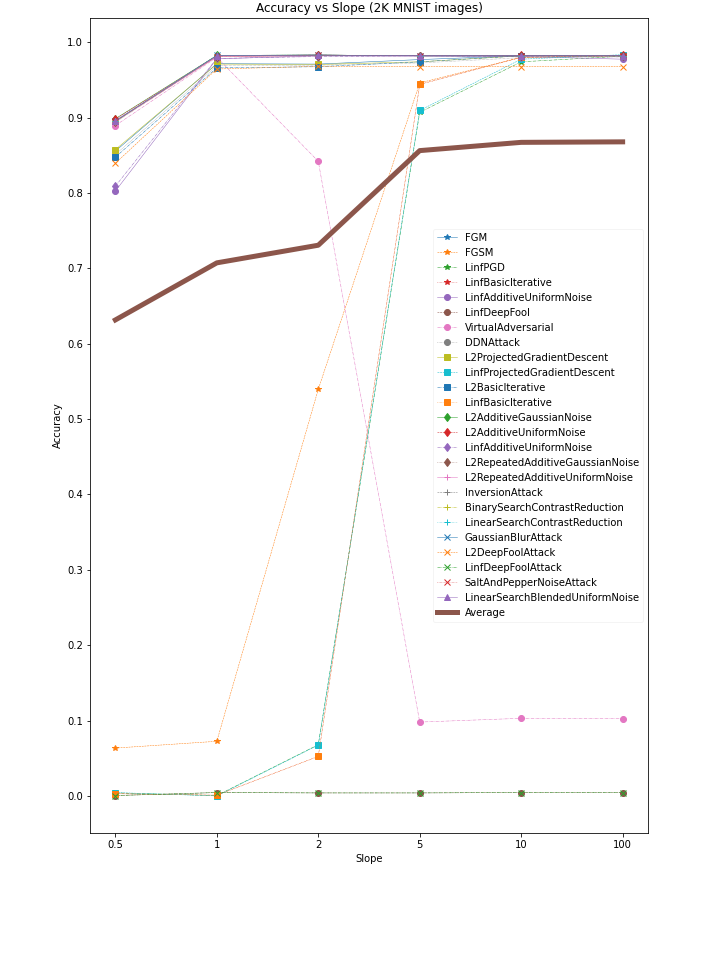}
    \hspace{-15pt}
    \includegraphics[width=.4\linewidth]{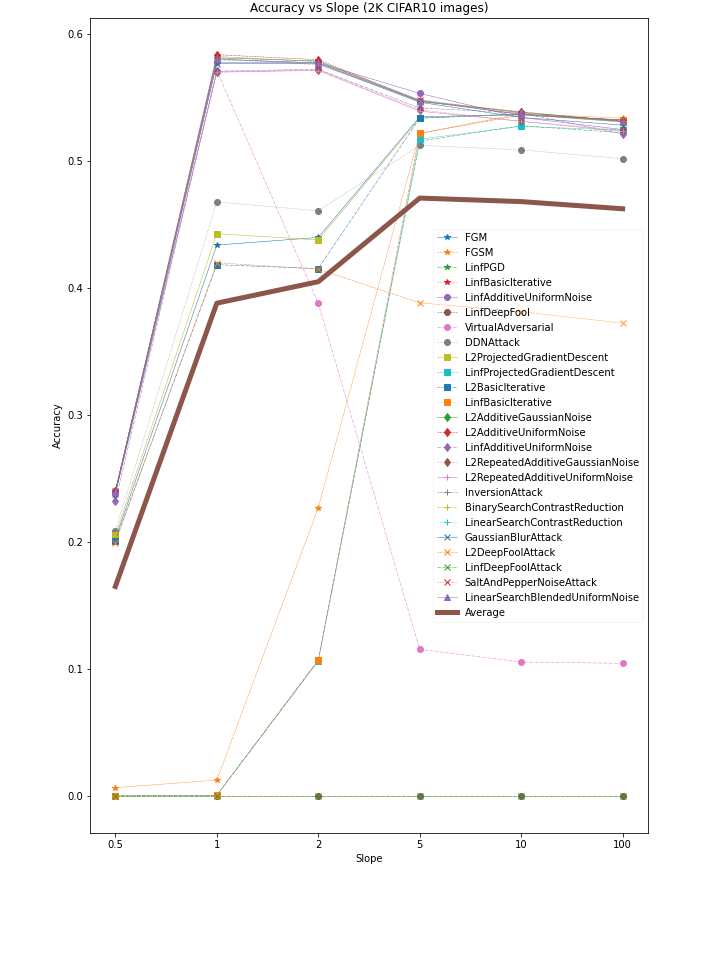}
    \hspace{-15pt}
    \includegraphics[width=.4\linewidth]{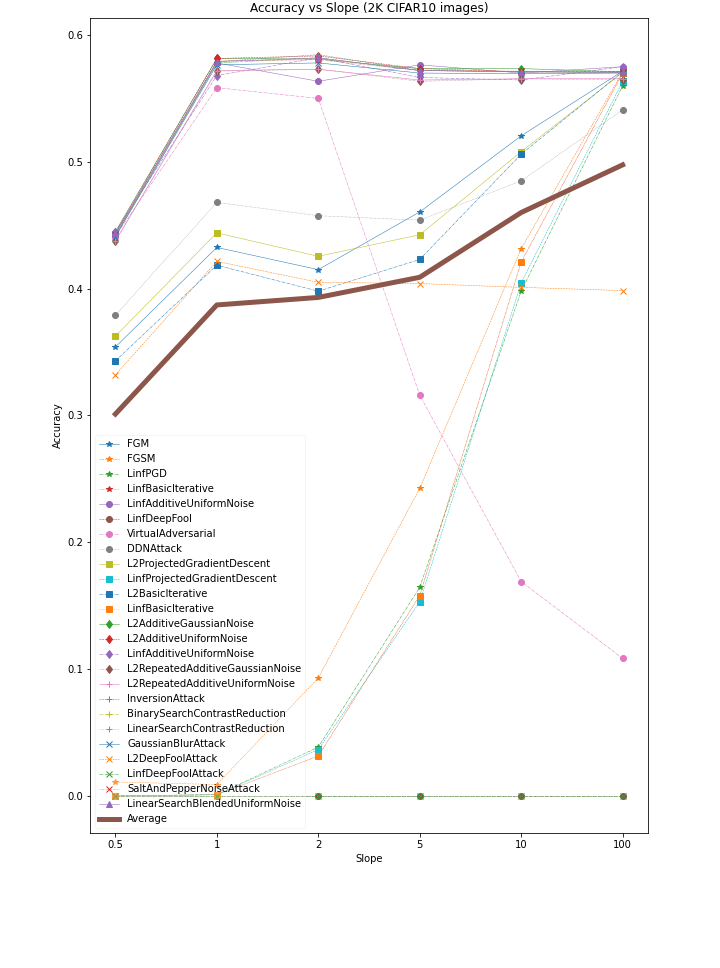}
        
    \caption{Results of ReLU defense against 25 untargeted attacks using the Foolbox code repository, over MNIST (left column), CIFAR10-CNN1 (middle), and CIFAR10-CNN2 (right). The first 2K images from each dataset were used.
    Top row shows the average performance over all attacks. Bottom row shows the performance at the last epsilon ($\epsilon = 0.3$ over MNIST and $\epsilon=0.1$ over CIFAR-10), for each attack. {\bf Higher y value here means better defense.}}
    \label{fig:foolbox}    
\end{adjustwidth}
\end{figure}

To assess this defense against a wider range of adversarial attacks, I conducted a larger scale analysis using the Foolbox code repository~\cite{rauber2017foolbox} over the first 2K images of each dataset. The proposed defense was tested against 25 attacks utilizing different L norms or image transformations (\eg Gaussian blurring). The top row in Fig.~\ref{fig:foolbox} shows the mean classifier accuracy averaged over all attacks. Increasing $\alpha$ improved the defense, although there seems to be a trade-off between accuracy and robustness. Please compare the last $\epsilon$ vs. $\epsilon=0$ in the top row of Fig.~\ref{fig:foolbox}. Looking across the attacks (bottom row in Fig.~\ref{fig:foolbox} shown only for the last $\epsilon$; $0.3$ over MNIST and $0.1$ over CIFAR), however, reveals that the boost in improvement comes only from few attacks against which the defense has done a great job, in particular FGSM and its variants (\eg PGD). Against the VirtualAdversarial attack, the proposed defense made the situation much worse. Overall, it seems that this defense works better against gradient based attacks as opposed to image degradation ones such as additive noise, salt and pepper noise, and inversion attack. Nonetheless, the ReLU defense was able to recover performance loss around 15\% over MNIST and around 10-15\% over CIFAR-10 (averaged over attacks; compare $\alpha=100$ vs. $\alpha=1$ in the bottom row of Fig.~\ref{fig:foolbox}).


\begin{figure}[t]       
\begin{center}
    \includegraphics[width=.5\linewidth]{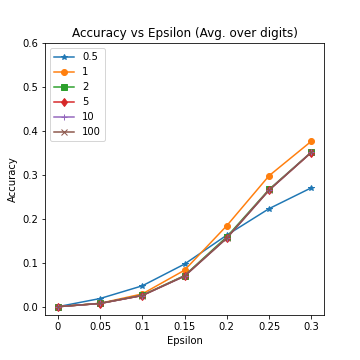} \\ 
    \vspace{-15pt}    
    \includegraphics[width=.5\linewidth]{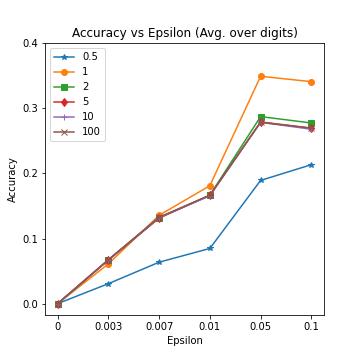} \\
    \vspace{-15pt}    
    \includegraphics[width=.5\linewidth]{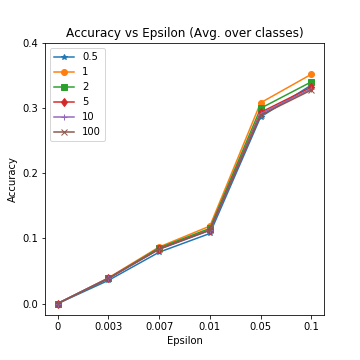}    
    \caption{Performance of the ReLU defense against the {\bf targeted FGSM attack}
    over MNIST (top row), CIFAR10-CNN1 (middle row), and CIFAR10-CNN2 (bottom row). {\bf Lower y value here means better defense.}}
        
\label{fig:targeted}
\end{center}    
\end{figure}

\begin{figure}[h]      
\begin{adjustwidth}{-2cm}{-2cm}
\begin{center}
    \vspace{-30px}    
    \includegraphics[width=.2\linewidth]{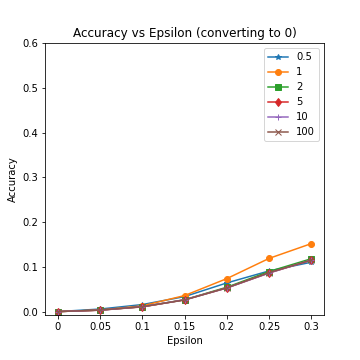}
    \hspace{-10px}
    \includegraphics[width=.2\linewidth]{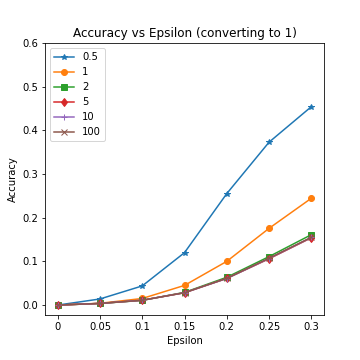}
    \hspace{-10px}
    \includegraphics[width=.2\linewidth]{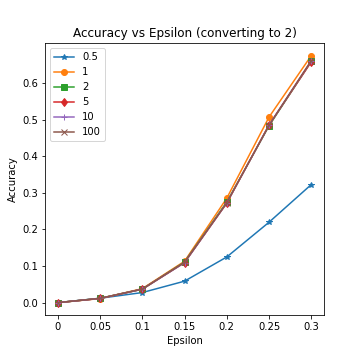} 
    \hspace{-10px}
    \includegraphics[width=.2\linewidth]{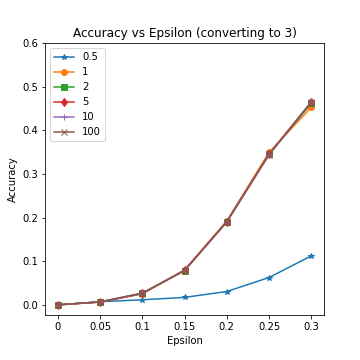} 
    \hspace{-10px}
    \includegraphics[width=.2\linewidth]{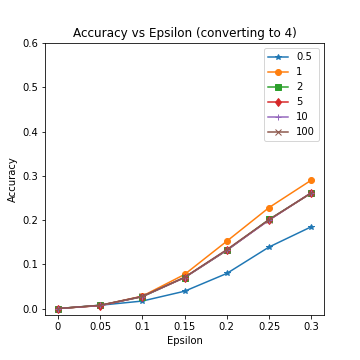} \\
    \includegraphics[width=.2\linewidth]{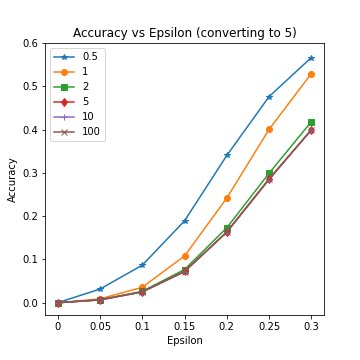} 
    \hspace{-10px}
    \includegraphics[width=.2\linewidth]{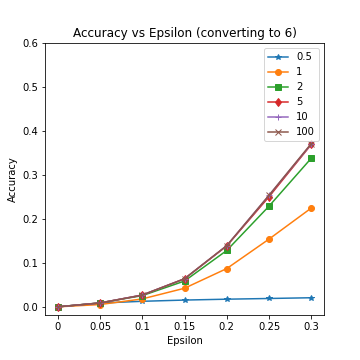}
    \hspace{-10px}
    \includegraphics[width=.2\linewidth]{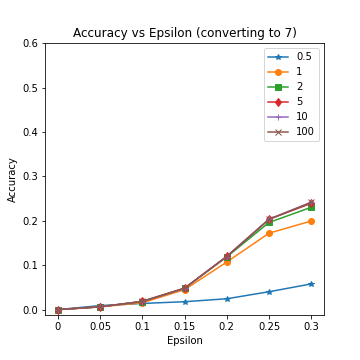} 
    \hspace{-10px}    
    \includegraphics[width=.2\linewidth]{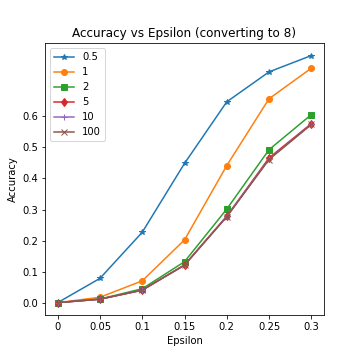} 
    \hspace{-10px}    
    \includegraphics[width=.2\linewidth]{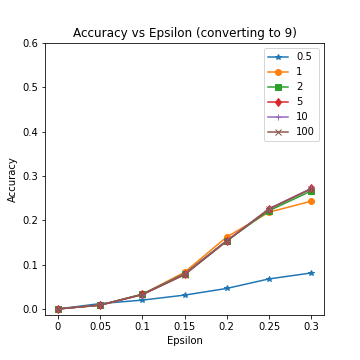}\\
    \line(1,0){480} \\
    \includegraphics[width=.2\linewidth]{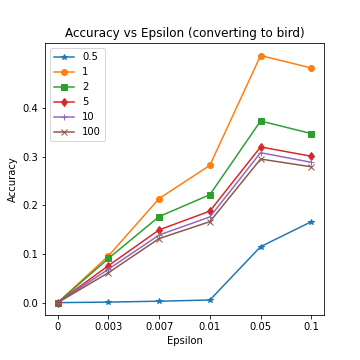}
    \hspace{-10px}
    \includegraphics[width=.2\linewidth]{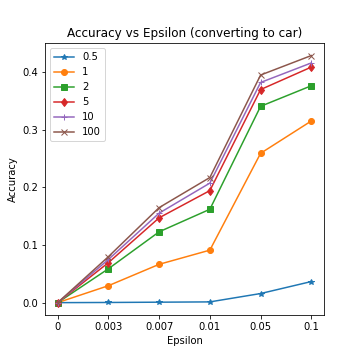}
    \hspace{-10px}
    \includegraphics[width=.2\linewidth]{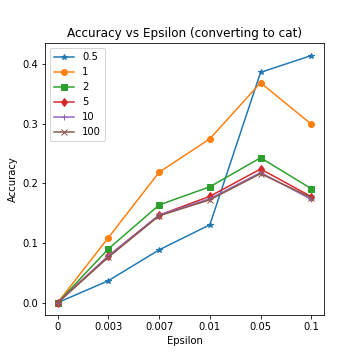} 
    \hspace{-10px}
    \includegraphics[width=.2\linewidth]{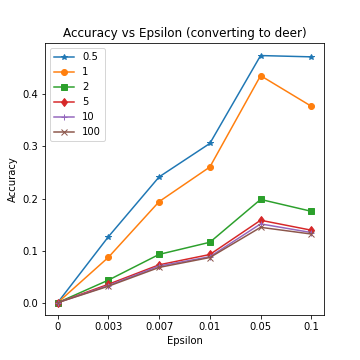}
    \hspace{-10px}
    \includegraphics[width=.2\linewidth]{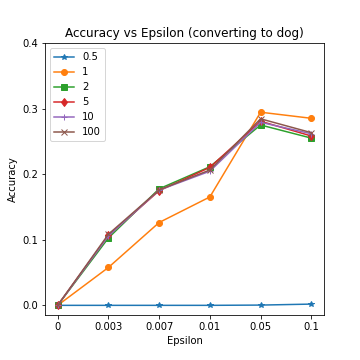} \\
    \includegraphics[width=.2\linewidth]{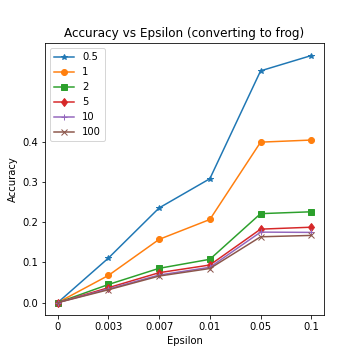} 
    \hspace{-10px}
    \includegraphics[width=.2\linewidth]{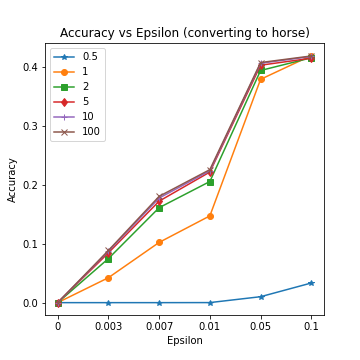}
    \hspace{-10px}
    \includegraphics[width=.2\linewidth]{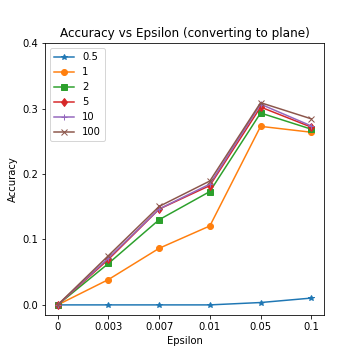} 
    \hspace{-10px}    
    \includegraphics[width=.2\linewidth]{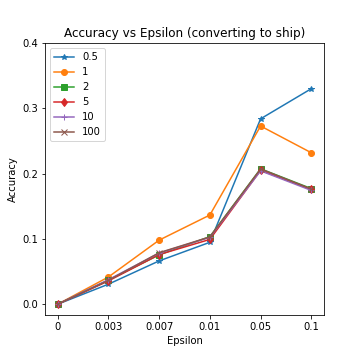} 
    \hspace{-10px}    
    \includegraphics[width=.2\linewidth]{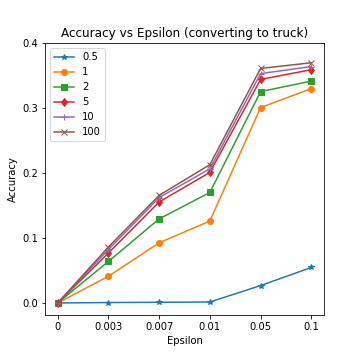}
    \line(1,0){480} \\
    \includegraphics[width=.2\linewidth]{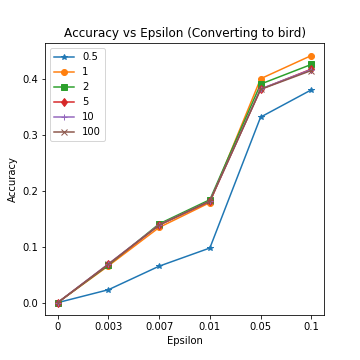}
    \hspace{-10px}
    \includegraphics[width=.2\linewidth]{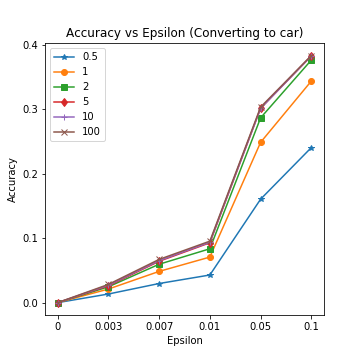}
    \hspace{-10px}
    \includegraphics[width=.2\linewidth]{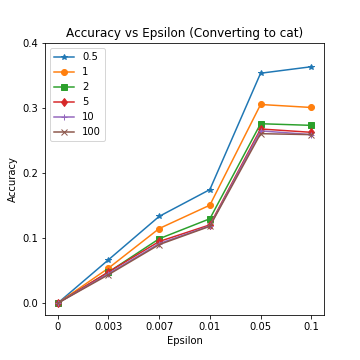} 
    \hspace{-10px}
    \includegraphics[width=.2\linewidth]{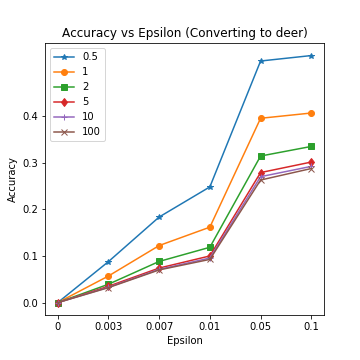} 
    \hspace{-10px}
    \includegraphics[width=.2\linewidth]{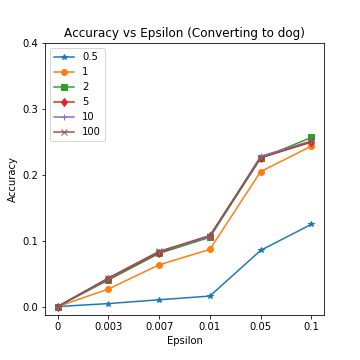} \\
    \includegraphics[width=.2\linewidth]{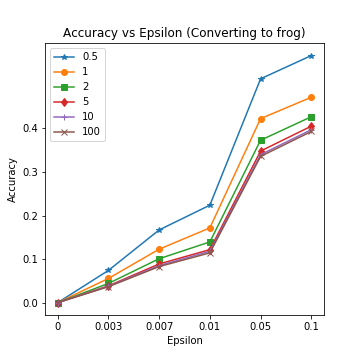} 
    \hspace{-10px}
    \includegraphics[width=.2\linewidth]{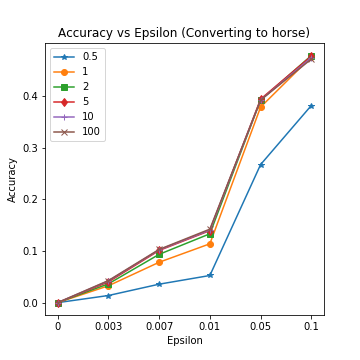}
    \hspace{-10px}
    \includegraphics[width=.2\linewidth]{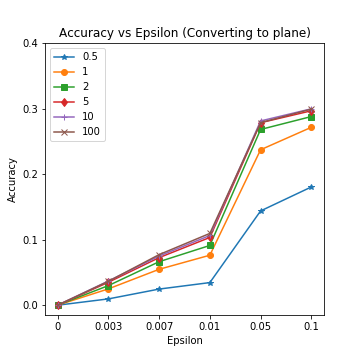} 
    \hspace{-10px}    
    \includegraphics[width=.2\linewidth]{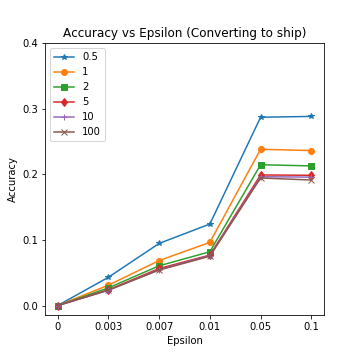} 
    \hspace{-10px}    
    \includegraphics[width=.2\linewidth]{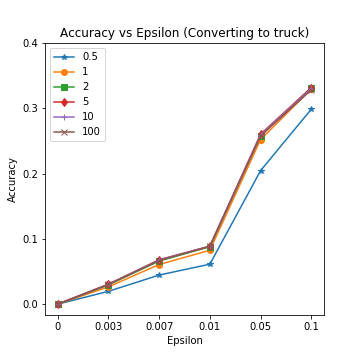}

    \caption{Performance of the ReLU defense against the {\bf targeted FGSM attack}
    over MNIST (top rows), CIFAR10-CNN1 (middle rows), and CIFAR10-CNN2 (bottom rows). Each plot corresponds to converting images to a different target class; if they have not already been classified as the target class). For some classes, the defense exacerbates the problem and helps the attack. On average, however, the defense seems to be working, although not as effective as in the untargeted setting. {\bf Lower y value here means better defense.}} 
    \label{fig:targeted_models}
\end{center}
\end{adjustwidth}
\end{figure}

\subsection{Performance against targeted attacks}
Performance of the ReLU defense against the targeted FGSM attack is reported in Fig.~\ref{fig:targeted} over the full test sets of MNIST and CIFAR-10 datasets. The y axis here shows the success rate of the attack in fooling the CNN to classify an image as the desired target class (if it has not already been classified so). As expected, increasing $\epsilon$ (perturbation magnitude) leads to higher attack accuracy. Over both datasets, increasing the SReLU slope (above 1) reduces the attack success rate which means better defense. Here, ReLU defense seems to perform about the same for slopes greater than one.

Defense results over individual classes are shown in Fig.~\ref{fig:targeted_models}. Increasing the slope helped in 6 classes, out of 10, using both MNIST CNN and CIFAR10-CNN1. Using CIFAR10-CNN2, 4 classes were defended better. In the remaining cases, the defense tied with the slope of 1 (\ie no defense) or slightly deteriorated the robustness. On average, it seems that ReLU defense is effective against the targeted FGSM attack, although not as good as its performance against the untargeted FGSM.

\section{Discussion and Conclusion}
What is happening? I suspect the reason why this defense works is because it enhances the signal-to-noise ratio. In the untargeted attack setting, increasing the loss leads to suppressing pixels that are important in making the correct prediction while boosting irrelevant features. Since relevant features are associated with higher weights, increasing the SReLU slope will enhance those feature more compared to the irrelevant ones, hence recovering good features. For the same reason, this defense is not very effective against targeted attacks. In this setting, the attacker aims to lower the loss in favor of the target class. Raising the SReLU slope results in enhancing features that are important to the target class, as well as the true class. This ignites a competition between two sets of features which can go either way. This explanation is reminiscent of attention mechanisms in human vision (and also in computer vision), in particular neural gain modulation theories. These theories explain behavioral and neural data in a variety of visual tasks such as discrimination and visual search~\cite{scolari2010basing,borji2014optimal}. On a related note, elevating the bias of neurons in different layers, as is done in ~\cite{borji2019white}, may lead to similar observations.

\noindent {\bf Gradient obfuscation.} An alternative, and perhaps more plausible, explanation is gradient explosion. Increasing the SReLU slope causes the gradients to explode during back-propagation. Strangely, however, this does not stop the targeted attacks! In this regard, the ReLU defense falls under the category of defenses that aim to mask or obfuscate gradients (\eg~\cite{papernot2017practical,xie2017mitigating,song2017pixeldefend,dhillon2018stochastic}). Some works have shown that these methods give a false sense of security and are thus breakable~\cite{athalye2018obfuscated}. Training and testing the network with the same slope (greater than one) did not improve the robustness, which could strengthen the gradient explosion reasoning.

\begin{figure}[htbp]       
\begin{center}
    \includegraphics[width=.4\linewidth]{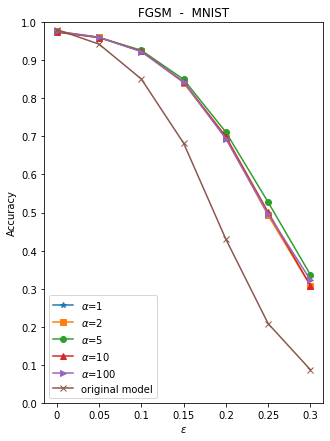} 
    \includegraphics[width=.4\linewidth]{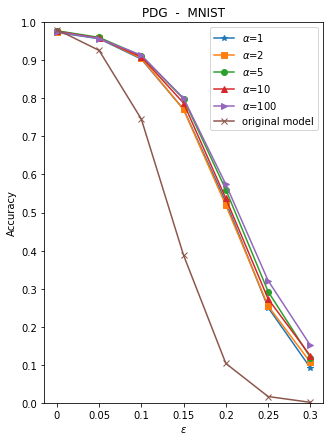} \\
    \includegraphics[width=.4\linewidth]{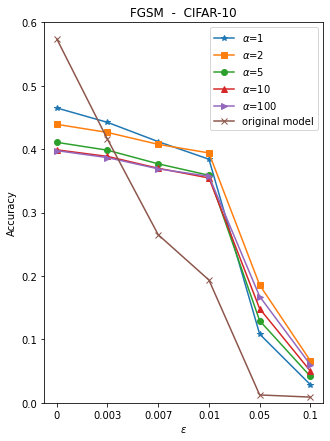}
    \includegraphics[width=.4\linewidth]{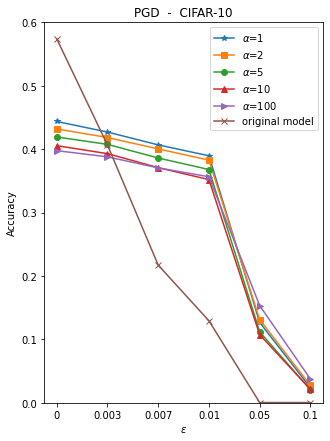}    
        
    \caption{Results of the BPDA attack (untargeted attack). {\bf Lower y value here means better defense.}}
        
\label{fig:BPDA}
\end{center}    
\end{figure}

To further investigate gradient obfuscation, I followed the approach proposed in~\cite{athalye2018obfuscated} known as Backward Pass Differentiable Approximation (BPDA). A second CNN, with the same architecture used in experiments, was trained. Instead of the cross entropy loss over predictions and ground truth labels, the model was trained with the cross entropy over logits of the two networks. The adversarial examples crafted for the second network were then applied to the first one. Results are shown in Fig.~\ref{fig:BPDA}. As it can be seen, the original models are severely hindered by the transfer attack using both FGSM and PGD attacks. The accuracy, however, is still higher than accuracy using attacks directly on the original models.

\begin{figure}[t]       
\begin{center}
\includegraphics[width=.385\linewidth]{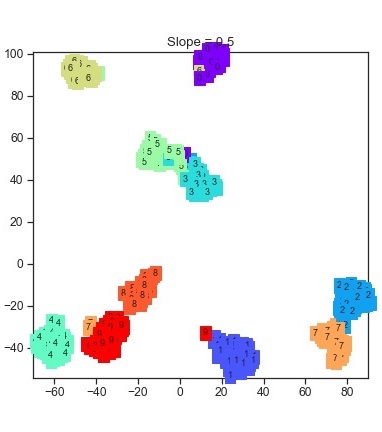} 
\includegraphics[width=.39\linewidth]{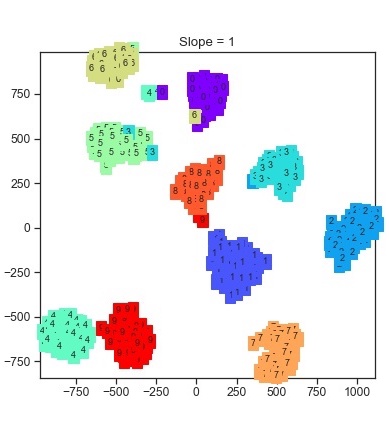} \\ \hspace{-5pt}
\includegraphics[width=.4\linewidth]{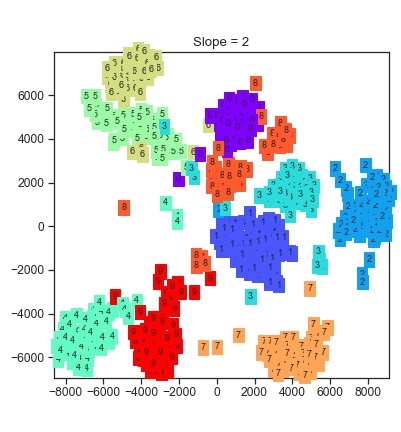} \includegraphics[width=.4\linewidth]{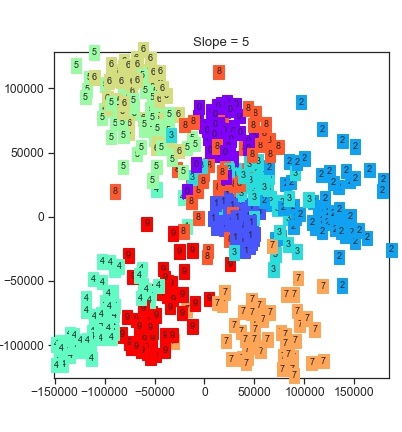}\\ \hspace{-5pt}
\includegraphics[width=.4\linewidth]{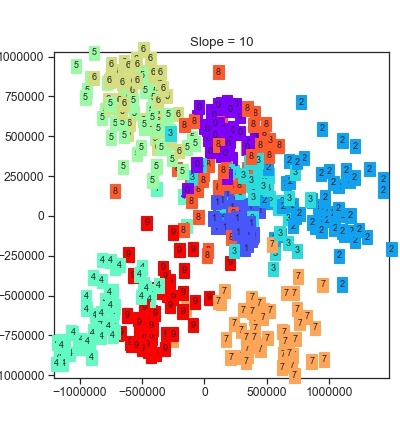} 
\includegraphics[width=.4\linewidth]{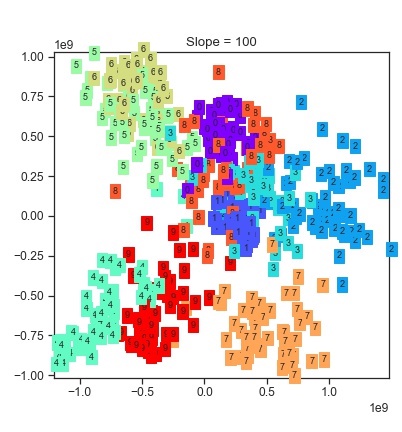} 
\caption{tSNE visualization of the last fully connected layer of the CNN trained over MNIST.}
\label{fig:tSNE}
\end{center}
\end{figure}

Fig.~\ref{fig:tSNE} shows tSNE illustration of the last fully connected layer over the MNIST dataset for different ReLU slopes. As the slope increases, the clusters become less compact. However, the distance between digits increases because of the higher magnitude of neuron activations due to the higher slopes.

\begin{figure}[h]       
\begin{center}
\includegraphics[width=.4\linewidth]{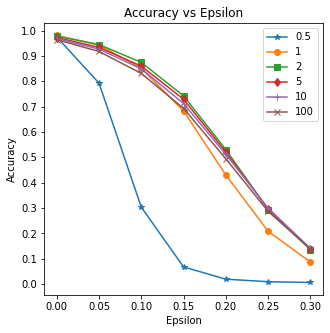} 
\includegraphics[width=.4\linewidth]{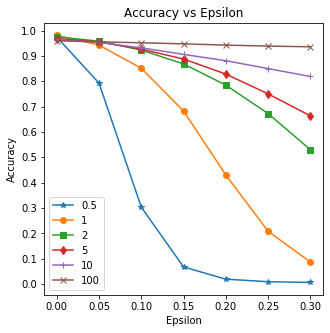} 
\includegraphics[width=.4\linewidth]{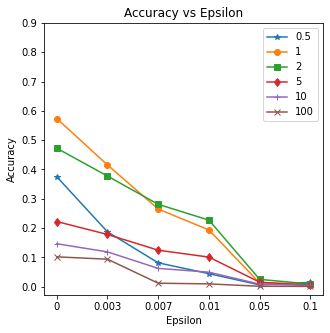} 
\includegraphics[width=.4\linewidth]{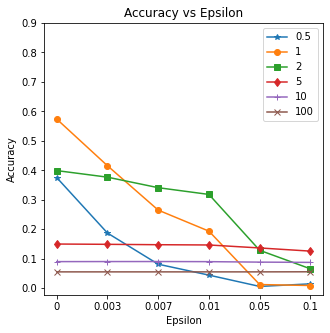} 
\caption{Impact of scaling up pixel values against the untargeted FGSM attak over MNIST (top) and CIFAR-10 (bottom; using CIFAR10-CNN1) datasets. The left column shows results with clipping the pixel values to [0,1] range after scaling. The right column shows results without clipping. The legend represents the magnitude of scaling. {\bf Higher y value here means better defense.}}
\label{fig:scale}
\end{center}
\end{figure}

\noindent {\bf Is increasing the SReLU slope the same as scaling up the image by a factor?} In general, No. Notice that $\alpha \text{max}(0,x) = \text{max}(0,\alpha x)$, for $\alpha \geq 0$ and $x \geq 0$. For a linear network with positive weights, the answer is yes, but for non-linear CNNs comprised of positive and negative weights the answer is no. Just to make sure, I ran an experiment in which I multiplied the pixel values by $\alpha$ and computed the classifier performance under the untargeted FGSM attack. Results are shown in Fig.~\ref{fig:scale}. Clipping the pixel values to the [0,1] range (after scaling) did not improve the robustness. Interestingly, without clipping the pixel values, results resemble those obtained by increasing the SReLU slope (top right panel in Fig.~\ref{fig:scale}; compare with Fig.~\ref{fig:untargeted}.). This suggests that maybe instead of increasing the SReLU slope we can just scale the pixel values! Results over the CIFAR-10 dataset, however, do not support this hypothesis. On this dataset, scaling does not help robustness with or without clipping. The discrepancy of results over two datasets can be attributed to the fact that MNIST digits are gray level whereas CIFAR-10 images are RGB. Increasing the pixel intensity of MNIST digits leads to maintaining high classifier accuracy while at the same time making the job of the attacker harder since now he has to increase the magnitude of the perturbation. In conclusion, this analysis suggests that increasing pixel values is not as effective as increasing the SReLU slope. This, however, needs further investigation. If the opposite is true, then it will have the following unfortunate consequence. To counter the ReLU defense, the attacker can simply submit the scaled down version of the input image.

Last but not least, that fact that this simple defense consistently works against some strong attacks (\eg PGD) is surprising. But please take these findings with a healthy dose of scepticism as I am still investigating them. A similar work has also been reported in~\cite{Erichson2019}.
Future work should evaluate the proposed approach on other models (\eg ResNet), datasets (\eg ImageNet), and against black-box attacks. Further, saliency and interpretability tools such as~\cite{Selvaraju_2017_ICCV} can be used to explain the observed phenomenon.

\vspace{10pt}
\noindent {\bf Acknowledgement}. I would like to thank Google for making the Colaboratory platform available.

{\small
\bibliographystyle{unsrtnat}
\bibliography{refs}
}

\end{document}